\documentclass[uplatex]{article}

\usepackage[dvipdfmx]{graphicx}

\usepackage[numbers]{natbib}
\usepackage{amsthm,amssymb,tikz-cd}

\usepackage{wrapfig}
\graphicspath{{../Picture1/}}

\makeatletter
\newcommand{\figcaption}[1]{\def\@captype{figure}\caption{#1}}
\newcommand{\tblcaption}[1]{\def\@captype{table}\caption{#1}}
\makeatother

\usepackage{comment} 

\usepackage[top=30truemm,bottom=30truemm,left=20truemm,right=20truemm]{geometry}
\usepackage{fancyhdr}
\usepackage{multirow}
\usepackage{here}
\usepackage{scalefnt}
\usepackage{indentfirst}
\usepackage{amsmath}
\usepackage{amssymb}
\usepackage{mathcomp}
\usepackage{mathtools}
\usepackage{cases}
\usepackage{cancel}
\usepackage{empheq}

\makeatother

\usepackage{authblk}
\author[1,*]{Shoki Iwaguchi}
\affil[1]{\footnotesize Department of Physics, Nagoya University, Nagoya, Aichi, 464-8602, Japan}
\author[2]{Atsushi Nishizawa}
\affil[2]{ Research Center for the Early Universe (RESCEU),
Graduate School of Science, The University of Tokyo, Tokyo 113-0033, Japan}
\author[3]{Yanbei Chen}
\affil[3]{Theoretical Astrophysics 350-17, California Institute of Technology, Pasadena, California 91125, USA}
\author[1]{Yuki Kawasaki}
\author[1]{Tomohiro Ishikawa}
\author[4,1]{Masaaki Kitaguchi}
\affil[4]{The Kobayashi-Maskawa Institute for the Origin of Particles and the Universe, Nagoya University, Nagoya, Aichi 464-8602, Japan}
\author[5]{Yutaka Yamagata}
\affil[5]{RIKEN Center for Advance Photonics, RIKEN, Hirosawa 2-1, Wako, Saitama 351-0198, JAPAN}
\author[1]{Bin Wu}
\author[1]{Ryuma Shimizu}
\author[1]{Kurumi Umemura}
\author[1]{Kenji Tsuji}
\author[1,4]{Hirohiko Shimizu}
\author[3]{Yuta Michimura}
\author[1,4]{Seiji Kawamura}

\affil[*]{Correspondence: iwaguchi\_s@u.phys.nagoya-u.ac.jp; +81-052-789-5982}

\title{Displacement-noise-free interferometeric gravitational-wave detector using unidirectional neutrons with four speeds}

\date{}

\begin{document}



\maketitle
\begin{abstract}
\noindent
For further gravitational wave (GW) detections, it is significant to invent a technique to reduce all kinds of mirror displacement noise dominant at low frequencies for ground-based detectors. The neutron displacement-noise-free interferometer (DFI) is one of the tools to reduce all the mirror displacement noise at lower frequencies. In this paper, we describe a further simplified configuration of a neutron DFI in terms of neutron incidence direction. In the new configuration, neutrons enter the interferometer with unidirectional incidence at four speeds as opposed to two bidirectional incidences of opposite directions at two speeds as reported previously. This simplification of the neutron DFI is significant for proof-of-principle experiments.
\\
\\
\noindent
{\em keywords:}
Gravitational wave ; Neutron interferometer ; Displacement-noise free interferometer ; Mach-Zehnder interferometer
\end{abstract}

\section{Introduction}
For gravitational wave (GW) detection, technical and theoretical ideas for reducing various noise sources have been invented and installed in the GW detectors \cite{LIGO_Tech}\cite{Virgo_Tech}. Due to these efforts, ground-based detectors such as LIGO and Virgo have detected GWs from BH-BH \cite{LIGO}, NS-NS \cite{Virgo}, and NS-BH binary systems \cite{LIGO and Virgo}. However, sensitivity in the low-frequency regime, below 10-20 Hz, has been limited by various sources of displacement noise. In this frequency band, there are significant science targets such as primordial GWs \cite{Kuroyanagi}. However, displacement noise is dominant at lower frequencies and this noise prevents GW detection in this frequency band. One of the solutions for this noise is space-based detectors such as DECIGO \cite{DECIGO_1}\cite{DECIGO_2}, LISA \cite{LISA}, and BBO \cite{BBO}. In space, suspension thermal noise and seismic noise can be removed because the test masses are not suspended. However, it will cost an enormous amount of money and time because of difficulties associated with space missions. In addition, test masses still have mirror thermal noise and radiation pressure noise even in space. Therefore, it is significant to invent a technique to reduce all kinds of mirror displacement noise dominant at low frequencies for ground-based detectors.

The displacement-noise-free interferometer (DFI) is one of the methods to cancel all the mirror displacement noise \cite{DFI_3}. The principle of the DFI is based on general relativity; GW signals can be distinguished from displacement noise in a coordinate in the transverse-traceless gauge \cite{DFI_1}\cite{DFI_2}. The DFI can remove mirror displacement noise without canceling the GW signal \cite{DFI_4}\cite{DFI_5}, at frequencies above which the GW period is comparable to the propagation time of laser light. However, the DFI effective frequency with km-scale arm lengths is on order of 100kHz, which is much higher than the low-frequency band (0.1Hz to 10Hz) where displacement noise is dominant. This is because the speed of light is too fast; the light propagation time in the km-scale DFI is much shorter than the GW period even at low frequencies. To address this problem, a DFI with neutrons, which propagates much more slowly than light, was proposed in \cite{DFNI_Nishizawa}. In a DFI with neutrons whose propagation time is comparable to the GW period at low frequencies, the DFI signal is sensitive to GWs. For example, the effective frequency of the neutron DFI can be reduced to 1Hz with km-scale arm length and km/sec-class neutron speed.

Another advantage of using neutrons is the capability of choosing the speed of neutrons freely as opposed to light. This advantage can lead to simplification of the DFI configuration. Whereas the original neutron DFI configuration is composed of two large and two small Mach-Zehnder interferometers (MZIs) with single-speed neutrons \cite{DFNI_Nishizawa}, the simplified neutron DFI configuration is composed of a single MZI with the fast and slow neutrons \cite{DFNI_Iwaguchi}. The number of MZIs is decreased by increasing the variety of neutron speeds. Simplifying the neutron DFI configuration is significant for GW detection, since in general a simpler detector has fewer practical obstacles in experiment and observation.

In this paper, we further simplify the configuration of a neutron DFI by modifying neutron incidence directions. In the new configuration, neutrons enter the interferometer with a single directional incidence as opposed to two bidirectional incidences of opposite directions in the previous configuration. The neutron source for the single directional incidence can be prepared more easily than the two incidences of opposite directions for the following two reasons. First, it is not easy to prepare two neutron beams of opposite directions from a single neutron source, because a change in a propagation direction of a neutron beam by reflection is, in general, very small. Second, simultaneous operation of two neutron sources with opposite beam directions is not usually available. For these reasons, in this paper we focus on simplification of the neutron incidence direction in neutron DFIs. In addition, this simplification can be applied to a proof-of-principle experiment at demonstration of a neutron DFI. This is because a neutron interferometer with a single directional incidence is a fundamental and essential scheme used for neutron optics. Although a proof-of-principle experiment for a laser DFI was performed in previous research \cite{DFI_5}, that of a neutron DFI has not yet been attempted. Therefore, simplification of the neutron incidence direction is significant for a proof-of-principle experiment for a neutron DFI.

We review and introduce the neutron DFI configurations in Section \ref{subsec:2.1}. Then we explain cancellation of mirror displacement noise with the combination of different-speed neutrons entering from one direction in Section \ref{subsec:2.2}, and cancellation of the beam splitter displacement noise and the constraints on the neutron speed in Section \ref{subsec:2.3}. We provide the GW response in the neutron DFI in Section \ref{sec:3}. Finally we present the discussion and conclusions in Section \ref{sec:4}.

\section{Neutron DFI using unidirectional neutrons without gravity}

\subsection{Configurations of neutron DFIs}
\label{subsec:2.1}

A series of neutron DFI configurations is shown in Fig 1.

\begin{figure}[h]
   \centering
   \includegraphics[clip,height=5.5cm]{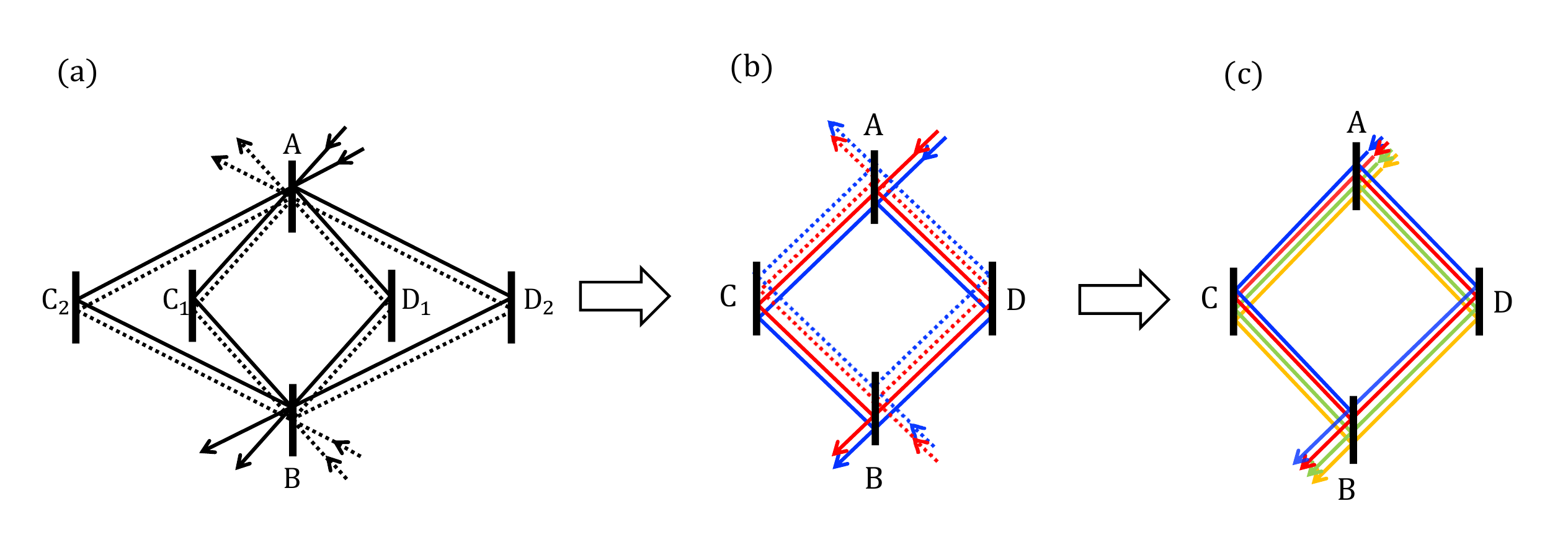}
   \caption{Configurations of neutron DFIs.}
   \label{fig:1}
\end{figure}

In Fig 1 (a), the original neutron DFI configuration is shown \cite{DFNI_Nishizawa}. Each MZI consists of a pair of counter-propagating neutrons beams. In this paper, this pair of counter-propagating neutrons is called a "bidirectional neutron." In this configuration, two bidirectional neutrons with a single speed enter two MZIs. Concerning the simplification of the DFI configuration, configuration (a) was modified to configuration (b). In Fig 1 (b), two bidirectional neutrons with two different speeds enter a single MZI \cite{DFNI_Iwaguchi}. Regarding the simplification of neutron incidence direction, configuration (b) has been modified to configuration (c). In Fig 1 (c), the new configuration proposed in this paper is shown. Neutrons with four different speeds enter a single MZI from a single port. In this paper, this neutron entering from one direction is called a "unidirectional neutron." Compared with previous research, experimental feasibility is increased because neutron source preparation is easier.

\subsection{Cancellation of mirror displacement noise by different-speed neutrons in the time domain}
\label{subsec:2.2}

In Fig 1 (c), four unidirectional neutrons with speeds $v_1, v_2, v_3,$ and $v_4$ ($v_1 > v_2 > v_3 > v_4$) enter a single MZI. The propagation time between the mirror and beam splitter is given by $T_\mathrm{i}$ (i=1,2,3,4). The displacements of the two mirrors and two beam splitters are defined as $x_{l} (t)$ $(l=\mathrm{A,B,C,D})$. When a neutron with speed $v_\mathrm{i}$ (i=1,2,3,4) hits the mirrors and beam splitters, the displacement noise is given by


\begin{equation}
   \phi_{l_\mathrm{i}} (t) = \kappa_\mathrm{i} x_{l} (t), \ \ \ \ \ \kappa_\mathrm{i} \equiv m v_\mathrm{i}.
   \label{eq:1}
\end{equation}

\noindent
Here, $m$ is the neutron rest mass and we use the unit of $c=1=\hbar$. We consider the case when they hit mirrors at the same time $t=t'$. When neutrons with the speed $v_\mathrm{i}$ (i=1,2,3,4) enter the MZIs, their interferometric signals $\phi_\mathrm{i}$ are given by

\begin{equation}
     \phi_\mathrm{i} (t') = \phi_{A_\mathrm{i}} (t'-T_\mathrm{i}) + \phi_{C_\mathrm{i}} (t') - \left \lbrace \phi_{D_\mathrm{i}} (t') + \phi_{B_\mathrm{i}} (t'+T_\mathrm{i}) \right \rbrace.
     \label{eq:2}
\end{equation}

\noindent
The displacement noise and impact times are denoted in Table \ref{tab:1}.

\begin{table}[h]
  \centering
  \caption{Displacement noise of mirrors and beam splitters in the case where neutrons hit mirrors at the same time $t=t'$. The subscripts denote the neutron speed and the impact point of the neutrons.}
  \scalebox{1}{
  {\renewcommand\arraystretch{2.0}
  \begin{tabular}{c||c|c|c|c} \hline
    Signal & C & D & A & B \\  \hline \hline
    $\phi_1(t)$ & $\phi_{\mathrm{C}_1} (t')$ & $\phi_{\mathrm{D}_1} (t')$ & $\phi_{\mathrm{A}_1} (t'-T_1)$ & $\phi_{\mathrm{B}_1} (t'+T_1)$ \\ \hline
    $\phi_2(t)$ & $\phi_{\mathrm{C}_2} (t')$ & $\phi_{\mathrm{D}_2} (t')$ & $\phi_{\mathrm{A}_2} (t'-T_2)$ & $\phi_{\mathrm{B}_2} (t'+T_2)$ \\ \hline
    $\phi_3(t)$ & $\phi_{\mathrm{C}_3} (t')$ & $\phi_{\mathrm{D}_3} (t')$ & $\phi_{\mathrm{A}_3} (t'-T_3)$ & $\phi_{\mathrm{B}_3} (t'+T_3)$ \\ \hline
    $\phi_4(t)$ & $\phi_{\mathrm{C}_4} (t')$ & $\phi_{\mathrm{D}_4} (t')$ & $\phi_{\mathrm{A}_4} (t'-T_4)$ & $\phi_{\mathrm{B}_4} (t'+T_4)$ \\  \hline
  \end{tabular}}
  }
  \label{tab:1}
\end{table}

The combinations of the signals required to cancel mirror displacement noise are given by

\begin{equation}
     \phi_{14} = \frac{1}{\kappa_1}\phi_{1} - \frac{1}{\kappa_4}\phi_{4},
     \label{eq:3}
\end{equation}

\begin{equation}
     \phi_{23} = \frac{1}{\kappa_2}\phi_{2} - \frac{1}{\kappa_3}\phi_{3}.
     \label{eq:4}
\end{equation}

The division by the speeds of $v_\mathrm{i}$ is the normalization of displacement noise for noise cancellation. Mirror noise cancellation is based on the condition that neutrons hit mirrors at the same time. Accordingly, the combination of the normalized signals of neutrons with different speeds can remove mirror displacement noise in the time domain.

\subsection{Cancellation of beam splitter displacement noise by unidirectional neutrons in the frequency domain}
\label{subsec:2.3}

Beam splitter displacement noise is present in signal combinations $\phi_{14}$ and $\phi_{23}$. For cancellation of beam splitter noise, we consider their noise in the frequency domain. In the frequency domain, displacement $x_{l} (t)$, an arbitrary term with an arbitrary frequency $\Omega'$, is written as

\begin{align}
    &  x_l(t) = \int X_{l}(\Omega) e^{i \lparen \Omega t + \varphi_{l}(\Omega) \rparen} d \Omega , \notag \\
   \therefore \ \ \ & x_{l}(\Omega') = X_{l}(\Omega') e^{i \varphi_{l}(\Omega')}.
   \label{eq:5}
\end{align}

\noindent
Here, $X_{l}(\Omega)$ and $\varphi_{l}(\Omega)$ are are the amplitude and the initial phase in the displacement. With Eq.(\ref{eq:1}), the displacement noise in the frequency domain is given by

\begin{align}
        \phi_{l_\mathrm{i}} (\Omega') = \kappa_{\mathrm{i}} X_{l}(\Omega') e^{i \varphi_{l}(\Omega')}.
        \label{eq:6}
\end{align}

\noindent
After the combination for mirror noise cancellation, the displacement noise of the beam splitters are shown in Table \ref{tab:2}.

\begin{table}[h]
 \centering
 \caption{Displacement noise of beam splitters after the combination to cancel mirror displacement noise.}
 \scalebox{1}{
  {\renewcommand\arraystretch{2.0}
  \begin{tabular}{c||c|c}
    \hline
    Signal & A & B \\
    \hline \hline
    $\phi_{14}$  &  $ x_\mathrm{A}(\Omega') \lbrace e^{-i \Omega' T_1} - e^{-i \Omega' T_4} \rbrace $  &  $ x_\mathrm{B}(\Omega') \lbrace e^{i \Omega' T_1} - e^{i \Omega' T_4} \rbrace $  \\ \hline
    $\phi_{23}$  &  $ x_\mathrm{A}(\Omega') \lbrace e^{-i \Omega' T_2} - e^{-i \Omega' T_3} \rbrace $  &  $ x_\mathrm{B}(\Omega') \lbrace e^{i \Omega' T_2} - e^{i \Omega' T_3} \rbrace $  \\ \hline
  \end{tabular}
  }
  }
  \label{tab:2}
\end{table}

We define the coefficients $c_{14}$ and $c_{23}$ for noise cancellation of beam splitters. With these coefficients, the displacement-noise-free combination $\phi_\mathrm{DFI}$ is given by

\begin{equation}
      \phi_\mathrm{DFI} = c_{14} \phi_{14} - c_{23} \phi_{23}.
      \label{eq:7}
\end{equation}

\noindent
Here, we add the condition for beam splitter noise cancellation, which is defined by

\begin{equation}
      T_1 + T_4 = T_2 + T_3.
      \label{eq:8}
\end{equation}

\noindent
With Eq.(\ref{eq:8}) and the displacement noise of the beam splitters shown in Table \ref{tab:2}, the coefficients $c_{14}$ and $c_{23}$ satisfy the equations given by

 \begin{align}
     & \ \ \ \  x_\mathrm{A}(\Omega') \lbrack c_{14} \lbrace e^{-i \Omega' T_1} - e^{-i \Omega' T_4} \rbrace - c_{23} \lbrace e^{-i \Omega' T_2} - e^{-i \Omega' T_3} \rbrace \rbrack \notag \\
     & = x_\mathrm{A}(\Omega') \lbrack c_{14} e^{-i \Omega' \frac{T_1+T_4}{2}} \lbrace e^{-i \Omega' \frac{T_1-T_4}{2}} - e^{i \Omega' \frac{T_1-T_4}{2}} \rbrace - c_{23} e^{-i \Omega' \frac{T_2+T_3}{2}} \lbrace e^{-i \Omega' \frac{T_2-T_3}{2}} - e^{i \Omega' \frac{T_2-T_3}{2}} \rbrace \rbrack \ \ \mathrm{and}  \label{eq:9}
 \end{align}

 \begin{align}
     & \ \ \ \ x_\mathrm{B}(\Omega') \lbrack c_{14} \lbrace e^{i \Omega' T_1} - e^{i \Omega' T_4} \rbrace - c_{23} \lbrace e^{i \Omega' T_2} - e^{i \Omega' T_3} \rbrace \rbrack \notag \\
     & = x_\mathrm{B}(\Omega') \lbrack c_{14} e^{i \Omega' \frac{T_1+T_4}{2}} \lbrace e^{i \Omega' \frac{T_1-T_4}{2}} - e^{-i \Omega' \frac{T_1-T_4}{2}} \rbrace - c_{23} e^{i \Omega' \frac{T_2+T_3}{2}} \lbrace e^{i \Omega' \frac{T_2-T_3}{2}} - e^{-i \Omega' \frac{T_2-T_3}{2}} \rbrace \rbrack. \label{eq:10}
 \end{align}

\noindent
Solving Eq.(\ref{eq:9})-(\ref{eq:10}) for $c_{14}$ and $c_{23}$ using Eq.(\ref{eq:8}), the coefficients $c_{14}$ and $c_{23}$ are given by

\begin{equation}
      c_{14} = \sin \Omega \left( \frac{T_2-T_3}{2} \right)  \ \ \mathrm{and} \ \ c_{23} = \sin \Omega \left( \frac{T_1-T_4}{2} \right).
      \label{eq:11}
\end{equation}

\noindent
Eq.(\ref{eq:7}) can be rewritten as

\begin{equation}
      \phi_\mathrm{DFI} = \sin \Omega \left( \frac{T_2-T_3}{2} \right) \times \phi_{14} - \sin \Omega \left( \frac{T_1-T_4}{2} \right) \times \phi_{23}.
      \label{eq:12}
\end{equation}

\begin{figure}[h]
   \centering
   \includegraphics[clip,height=10cm]{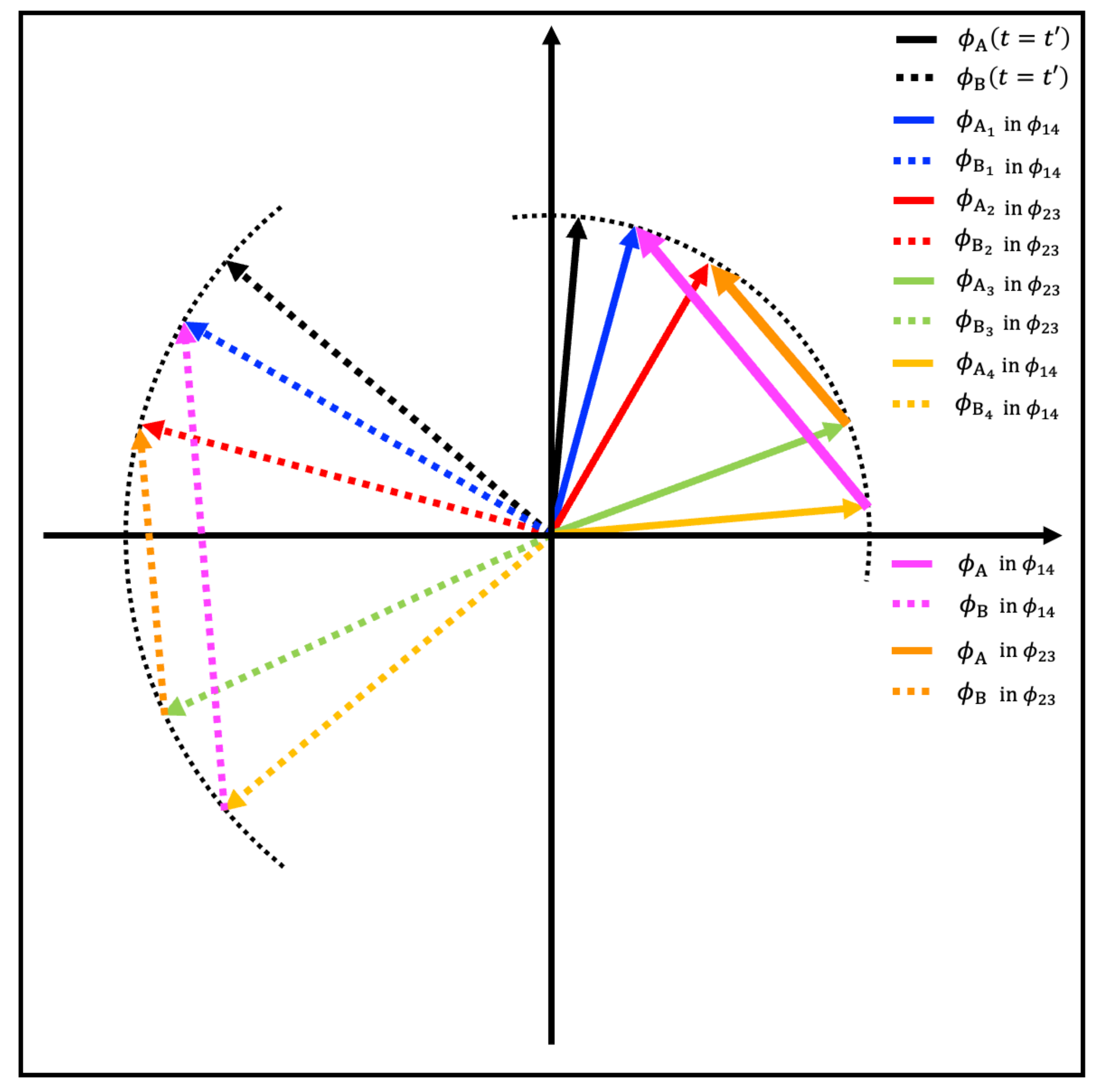}
   \caption{Phasor diagram for cancellation of beam splitter noise. The arrow length corresponds to the noise amplitude. The solid and dashed arrows show the displacement noise about the beam splitter A and B. The colors of the arrows show the neutron speeds. $v_1, v_2, v_3,$ and $v_4$ correspond to blue, red, green, and yellow respectively. Magenta and orange arrows show the displacement noise of the beam splitter in $\phi_{14}$ and $\phi_{23}$.}
   \label{fig:2}
\end{figure}

\noindent
The combination $\phi_\mathrm{DFI}$ has no mirror or beam splitter displacement noise. This mechanism can be explained by a phasor diagram, as shown in Fig.\ref{fig:2}. $\phi_{14}$ and $\phi_{23}$ have residual displacement noise from the beam splitter, which are defined as $\phi_{\mathrm{A}}$ and $\phi_{\mathrm{B}}$. They are represented by magenta and orange arrows in Fig.\ref{fig:2}. The condition given by Eq.(\ref{eq:8}) allows these arrows to be parallel. The difference between $\phi_{14}$ and $\phi_{23}$ is only the noise amplitude. As a result, the coefficient to equalize the amplitudes enables the cancellation of residual noise. Without this condition shown in Eq.(\ref{eq:10}), the magenta and orange arrows are not parallel, which leads to incomplete cancellation of the mirror displacement noise. Therefore, the condition of parallelization is essential for a neutron DFI using unidirectional neutrons.

\newpage

\section{Response to gravitational waves without gravity}
\label{sec:3}

\begin{figure}[h]
   \centering
   \includegraphics[clip,height=8cm]{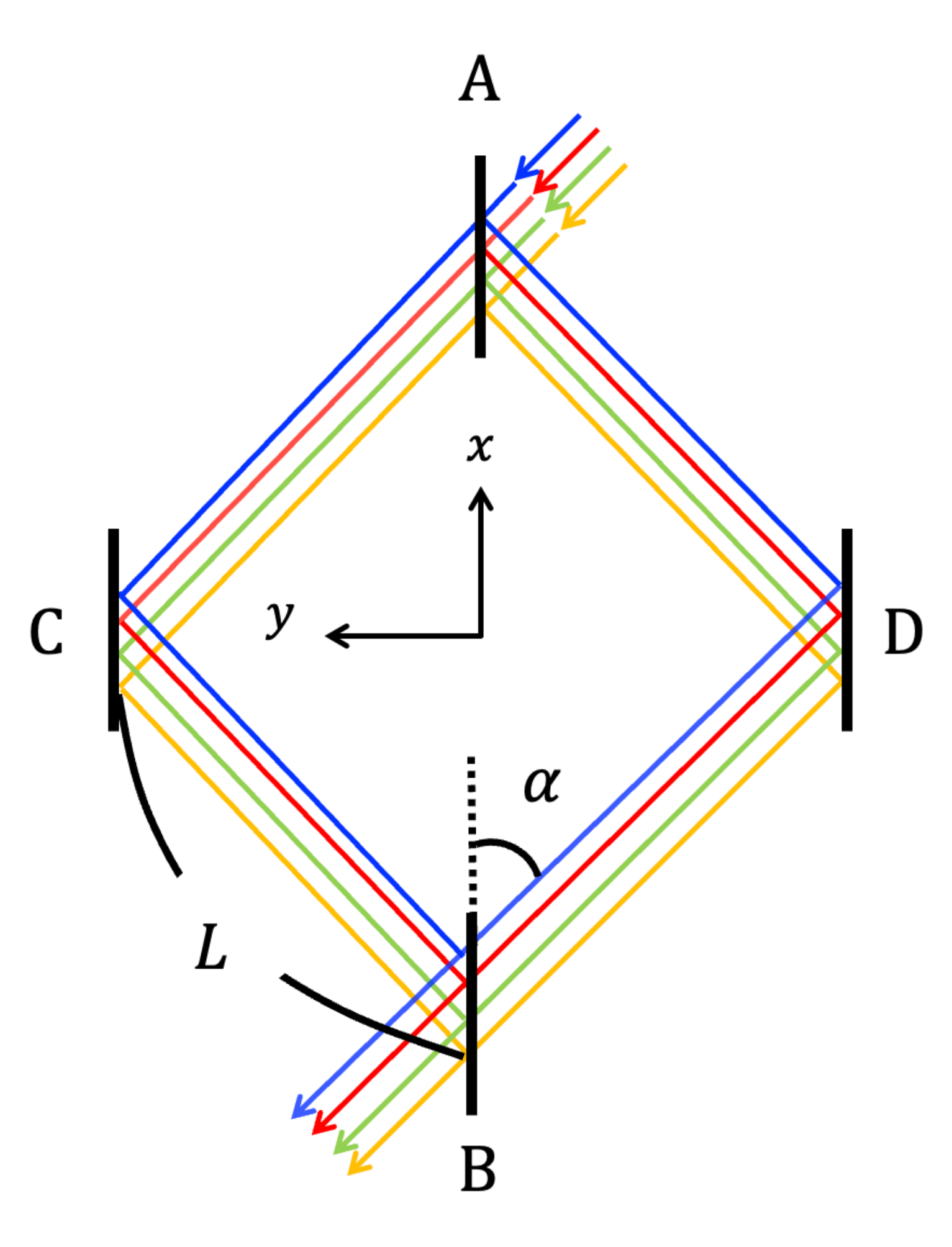}
   \caption{Definition of parameters.}
   \label{fig:3}
\end{figure}

In this section, we discuss the neutron DFI under the condition of the absence of gravity. The neutron DFI configuration and parameters are defined in Fig \ref{fig:3}. In this configuration, four unidirectional neutrons with speeds $v_1, v_2, v_3,$ and $v_4$ ($v_1 > v_2 > v_3 > v_4$) enter the single MZI.

\noindent
$T_\mathrm{i}$ is the transit time for each neutron between the beam splitters and the mirrors, which is given by

\begin{equation}
        T_\mathrm{i} = \frac{L}{v_{\mathrm{i}}}. \ \ \ \ \ (\mathrm{i} = 1,2,3,4),
        \label{eq:13}
\end{equation}

\noindent
The coordinates of the neutron incidence point on beamsplitter A are set as follows:

\begin{equation}
     \mathbf{x}_\mathrm{A} = \lbrace  L \cos \alpha , 0 ,0  \rbrace.
     \label{eq:14}
\end{equation}

\noindent
The neutron trajectories (A→C) are given by

\begin{equation}
        \mathbf{x}_\mathrm{{AC}_i} (t) = \lbrace - v_\mathrm{i} t \cos \alpha , v_\mathrm{i} t \sin \alpha , 0 \rbrace \ \ \ \ \ (0 \leq t \leq T_\mathrm{i}).
        \label{eq:15}
\end{equation}
\normalsize

\noindent
The coordinates of the neutron impact points on mirror C are given by
\begin{align}
     \mathbf{x}_\mathrm{C_i} &= \mathbf{x}_\mathrm{A} + \mathbf{x}_\mathrm{{AC}_i} (T_\mathrm{i}) \notag \\
                    &= \lbrace 0 , L \sin \alpha , 0 \rbrace.
                    \label{eq:16}
\end{align}

\noindent
The neutron trajectories (C→B) are given by

\begin{equation}
        \mathbf{x}_\mathrm{{CB}_i} (t) = \lbrace - v_\mathrm{i} t \cos \alpha , - v_\mathrm{i} t \sin \alpha , 0 \rbrace \ \ \ \ \ (0 \leq t \leq T_\mathrm{i}).
        \label{eq:17}
\end{equation}

\noindent
The coordinates of the neutron impact points on beam splitter B are given by
\begin{align}
     \mathbf{x}_\mathrm{B_i} &= \mathbf{x}_\mathrm{C} + \mathbf{x}_\mathrm{{CB}_i} (T_\mathrm{i}) \notag \\
                    &= \lbrace - L \cos \alpha, 0 ,0  \rbrace.
                    \label{eq:18}
\end{align}

\noindent
The neutron trajectories (A→D) are given by

\begin{equation}
        \mathbf{x}_\mathrm{{AD}_i} (t) = \lbrace - v_\mathrm{i} t \cos \alpha , - v_\mathrm{i} t \sin \alpha , 0 \rbrace \ \ \ \ \ (0 \leq t \leq T_\mathrm{i}).
        \label{eq:19}
\end{equation}
\normalsize

\noindent
The coordinates of the neutron impact points on mirror D are given by
\begin{align}
     \mathbf{x}_\mathrm{D_i} &= \mathbf{x}_\mathrm{A} + \mathbf{x}_\mathrm{{AD}_i} (T_\mathrm{i}) \notag \\
                    &= \lbrace 0 , - L \sin \alpha , 0 \rbrace.
                    \label{eq:20}
\end{align}

\noindent
The neutron trajectories (D→B) are given by

\begin{equation}
        \mathbf{x}_\mathrm{{DB}_i} (t) = \lbrace - v_\mathrm{i} t \cos \alpha , v_\mathrm{i} t \sin \alpha , 0 \rbrace \ \ \ \ \ (0 \leq t \leq T_\mathrm{i}).
        \label{eq:21}
\end{equation}

\noindent
The wavenumbers of the neutrons propagating from A to B are given by


\begin{align}
       & \mathbf{k}_\mathrm{{AC}_i} = m v_\mathrm{i} \lbrace - \cos \alpha , \sin \alpha , 0 \rbrace, \label{eq:22}\\
       & \mathbf{k}_\mathrm{{CB}_i} = m v_\mathrm{i} \lbrace - \cos \alpha , - \sin \alpha , 0 \rbrace, \label{eq:23}\\
       & \mathbf{k}_\mathrm{{AD}_i} = m v_\mathrm{i} \lbrace - \cos \alpha , - \sin \alpha , 0 \rbrace, \label{eq:24}\\
       & \mathbf{k}_\mathrm{{DB}_i} = m v_\mathrm{i} \lbrace - \cos \alpha , v_\mathrm{i} \sin \alpha , 0 \rbrace \label{eq:25}.
\end{align}


\noindent
The dimensionless wavenumbers of the neutrons are defined as $\tilde{\mathbf{k}} = \mathbf{k} / \kappa_\mathrm{i}$. The neutron's phase shift from GWs is $\phi_\mathrm{gw}$, which is derived from the Klein-Gordon equation \cite{DFNI_Nishizawa}. It is defined by

\begin{equation}
        \frac{\partial \phi_\mathrm{gw}}{\partial t} \approx - \frac{h^{ab} k_a k_b}{2 m}.
        \label{eq:26}
\end{equation}

\noindent
In this neutron DFI, when a neutron propagates from A at $t$ to C at $t+T_\mathrm{i}$ under the condition that they impact mirrors at the same time, the phase shift from GWs is given by

\begin{equation}
        \phi^\mathrm{gw}_\mathrm{{AC}_\mathrm{i}} (t) = -\frac{\kappa_\mathrm{i}^2}{2 m} \int_{t}^{t+T_\mathrm{i}} h^{ab} \lbrack t' , \mathbf{x}_\mathrm{AC} (t') \rbrack  \tilde{k}_{\mathrm{{AC}}_a} (t') \tilde{k}_{\mathrm{{AC}}_b} (t') dt'.
        \label{eq:27}
\end{equation}
\normalsize

\noindent
We define the timing noise $\phi^\mathrm{clock}_\mathrm{{AC}_\mathrm{i}} (t)$ and the displacement noise $\phi^\mathrm{disp}_\mathrm{{AC}_\mathrm{i}} (t)$ as

\begin{equation}
      \phi^\mathrm{clock}_\mathrm{{AC}_\mathrm{i}} (t) \approx m \lbrace \tau_\mathrm{C} (t+T_\mathrm{i}) - \tau_\mathrm{A} (t) \rbrace \ \ \mathrm{and}
      \label{eq:28}
\end{equation}

\begin{equation}
      \phi^\mathrm{disp}_\mathrm{{AC}_\mathrm{i}} (t) = \phi_{\mathrm{C}_\mathrm{i}} (t+T_\mathrm{i}) - \phi_{\mathrm{A}_\mathrm{i}} (t).
      \label{eq:29}
\end{equation}

\noindent
Here, $\tau_l$ is defined as the clock noise at location $l$ $(l = \mathrm{A, B, C, D})$. At the GW angular frequency $\Omega$, the Fourier transform of $h^{ab}$ is defined as

\begin{equation}
        H^{ab} (\Omega) \equiv \int_{- \infty}^{\infty} dt e^{i \Omega t} h^{ab} \lbrack t' , \mathbf{x} (t') \rbrack.
        \label{eq:30}
\end{equation}

\noindent
In the Fourier domain, the GW signal $\Phi^\mathrm{gw}_\mathrm{{AC}_i} (\Omega)$ is given by


\begin{align}
        \Phi^\mathrm{gw}_\mathrm{AC_i} (\Omega) &= - \frac{\kappa_\mathrm{i}^2}{2 m} \left \lbrace  P_0 (\Omega) \tilde{k}_{\mathrm{AC}_I} \tilde{k}_{\mathrm{AC}_J} H^{IJ} (\Omega) \right \rbrace,
        \label{eq:31}
\end{align}

\normalsize

\noindent
and the clock noise $\Phi^\mathrm{clock}_\mathrm{{AC}_\mathrm{i}} (\Omega)$ and the displacement noise $\Phi^\mathrm{disp}_\mathrm{{AC}_\mathrm{i}} (\Omega)$ are given by

\begin{equation}
      \Phi^\mathrm{clock}_{\mathrm{AC}_\mathrm{i}} (\Omega) \approx m \lbrace {\omega_\mathrm{i} (\Omega)} \tau_\mathrm{C} (\Omega) - \tau_\mathrm{A} (\Omega) \rbrace \ \mathrm{and}
      \label{eq:32}
\end{equation}

\begin{equation}
      \Phi^\mathrm{disp}_{\mathrm{AC}_\mathrm{i}} (\Omega) = {\omega_\mathrm{i} (\Omega)} \phi_{\mathrm{C}_\mathrm{i}} (\Omega) - \phi_{\mathrm{A}_\mathrm{i}} (\Omega).
      \label{eq:33}
\end{equation}

\noindent
Here, we define

\begin{equation}
        \omega_\mathrm{i} (\Omega) \equiv e^{- i \Omega T_\mathrm{i}} \ \ \ \mathrm{and}
        \label{eq:34}
\end{equation}

\begin{equation}
        P_0 (\Omega) \equiv -\frac{i}{\Omega} \lbrace 1 - \omega_\mathrm{i} (\Omega) \rbrace.
        \label{eq:35}
\end{equation}

\noindent
With Eq. (\ref{eq:31})-(\ref{eq:33}), in the Fourier domain, the signal for the neutron propagation from A to C is given  by


\begin{align}
        \Phi_\mathrm{AC_i} (\Omega) &= - \frac{\kappa_\mathrm{i}^2}{2 m} \left \lbrace P_0 (\Omega) \tilde{k}_{\mathrm{AC}_I} \tilde{k}_{\mathrm{AC}_J} H^{IJ} (\Omega) \right \rbrace + \Phi^\mathrm{clock}_\mathrm{{AC}_\mathrm{i}} (\Omega) + \Phi^\mathrm{disp}_\mathrm{{AC}_\mathrm{i}} (\Omega).
        \label{eq:36}
\end{align}

\normalsize

\noindent
The GW response function without the clock and displacement noise is given by


\begin{align}
      R_{\mathrm{AC_i}} (\Omega) \equiv \frac{|\Phi_\mathrm{AC_i} (\Omega)|}{\eta_\mathrm{i} H}  \ \ \ \ \mathrm{where} \ \ \eta_\mathrm{i} \equiv \frac{{\kappa_\mathrm{i}}^2 T_\mathrm{i}}{2 m}.
      \label{eq:37}
\end{align}

\normalsize

\noindent
Here, $H$ is the GW amplitude. Interferometric signals in this configuration are given by

\begin{equation}
       \phi_{\mathrm{AB}_\mathrm{i}} (t) = \phi_{\mathrm{AC_i}} (t) + \phi_{\mathrm{CA_i}} (t+T_\mathrm{i}) - \lbrace \phi_{\mathrm{AD_i}} (t) + \phi_{\mathrm{DB_i}} (t+T_\mathrm{i}) \rbrace .
       \label{eq:38}
\end{equation}

\noindent
In the Fourier domain, these interferometric signals can be written as

\begin{equation}
       \Phi_{\mathrm{AB}_\mathrm{i}} (\Omega) = \Phi_{\mathrm{AC_i}} (\Omega) + \omega_\mathrm{i} (\Omega) \Phi_{\mathrm{CB_i}} (\Omega) - \lbrace \Phi_{\mathrm{AD_i}} (\Omega) + \omega_\mathrm{i} (\Omega)  \Phi_{\mathrm{DB_i}} (\Omega) \rbrace .
       \label{eq:39}
\end{equation}

\noindent
From Eq.(\ref{eq:3})-(\ref{eq:4}), the signal combinations that cancel mirror displacement noise are given by

\begin{equation}
       \Phi_{14} (\Omega) = \Phi_{\mathrm{AB}_1} (\Omega) / \kappa_1 - \Phi_{\mathrm{AB}_4} (\Omega) / \kappa_4 \ \ \mathrm{and}
       \label{eq:40}
\end{equation}

\begin{equation}
       \Phi_{23} (\Omega) = \Phi_{\mathrm{AB}_2} (\Omega) / \kappa_2 - \Phi_{\mathrm{AB}_3} (\Omega) / \kappa_3 .
       \label{eq:41}
\end{equation}

\noindent
Consequently, using Eq.(\ref{eq:12}), the neutron DFI signal in the Fourier domain is given by

\begin{equation}
       \Phi_\mathrm{DFI} (\Omega) = \frac{1}{2 \Omega \bar{T}} \left \lbrace c_{14} (\Omega) \Phi_{14} (\Omega) -  c_{23} (\Omega) \Phi_{23} (\Omega) \right \rbrace \ \ \mathrm{where}
       \label{eq:42}
\end{equation}

\begin{align}
 c_{14} (\Omega) \equiv \frac{\sin \Omega \left( \frac{T_2-T_3}{2} \right)}{\sin \alpha}, \ \ \  c_{23} (\Omega) \equiv \frac{\sin \Omega \left( \frac{T_1-T_4}{2} \right)}{\sin \alpha} \ \ \mathrm{and} \ \ \bar{T} \equiv \frac{(T_2-T_3)/2+(T_1-T_4)/2}{2}.
      \label{eq:43}
\end{align}

\noindent
Here, the coefficients $c_{14} (\Omega)$ and $c_{23} (\Omega)$ are the frequency-dependent coefficients required to cancel displacement noise and normalization terms. The division by $2 \Omega \bar{T}$ plays a role in maintaining the neutron DFI response at lower frequencies. When a GW with a strain $h_{ij}$ and a polarization angle $\psi$ propagates from an arbitrary direction $(\phi,\theta)$, the rotation matrix is given by


\begin{equation}
    \mathcal{R} =
    \begin{pmatrix}
      \cos \phi & \sin \phi & 0 \\
      - \sin \phi & \cos \phi & 0 \\
      0 & 0 & 1
    \end{pmatrix}
    \begin{pmatrix}
      \cos \theta & 0 & - \sin \theta \\
      0 & 1 & 0 \\
      \sin \theta & 0 & \cos \theta
    \end{pmatrix}
    \begin{pmatrix}
      \cos \psi & \sin \psi & 0 \\
      - \sin \psi & \cos \psi & 0 \\
      0 & 0 & 1
    \end{pmatrix},
    \label{eq:44}
\end{equation}

\normalsize

\noindent
and the GW strain $h'_{ab}$ is written as

\begin{equation}
    h'_{ab} = \mathcal{R}_{ap} \mathcal{R}_{bq} h_{pq} = (\mathcal{R} h \mathcal{R}^T)_{ij}.
    \label{eq:45}
\end{equation}

\noindent
From Eq.(45), the DFI response function $R_\mathrm{DFI}$ is given by

\begin{equation}
   R_\mathrm{DFI} (\Omega) \equiv \frac{|\Phi_\mathrm{DFI} (\Omega)|}{\bar{\eta} H} \ \ \ \mathrm{where} \ \ \ {\bar{\eta}} \equiv \frac{\bar{\kappa} \bar{T}}{2 m} .
   \label{eq:46}
\end{equation}

\begin{figure}[h]
 \begin{tabular}{cc}
    \begin{minipage}[b]{0.45\hsize}
      \centering
      \includegraphics[clip,width=7cm]{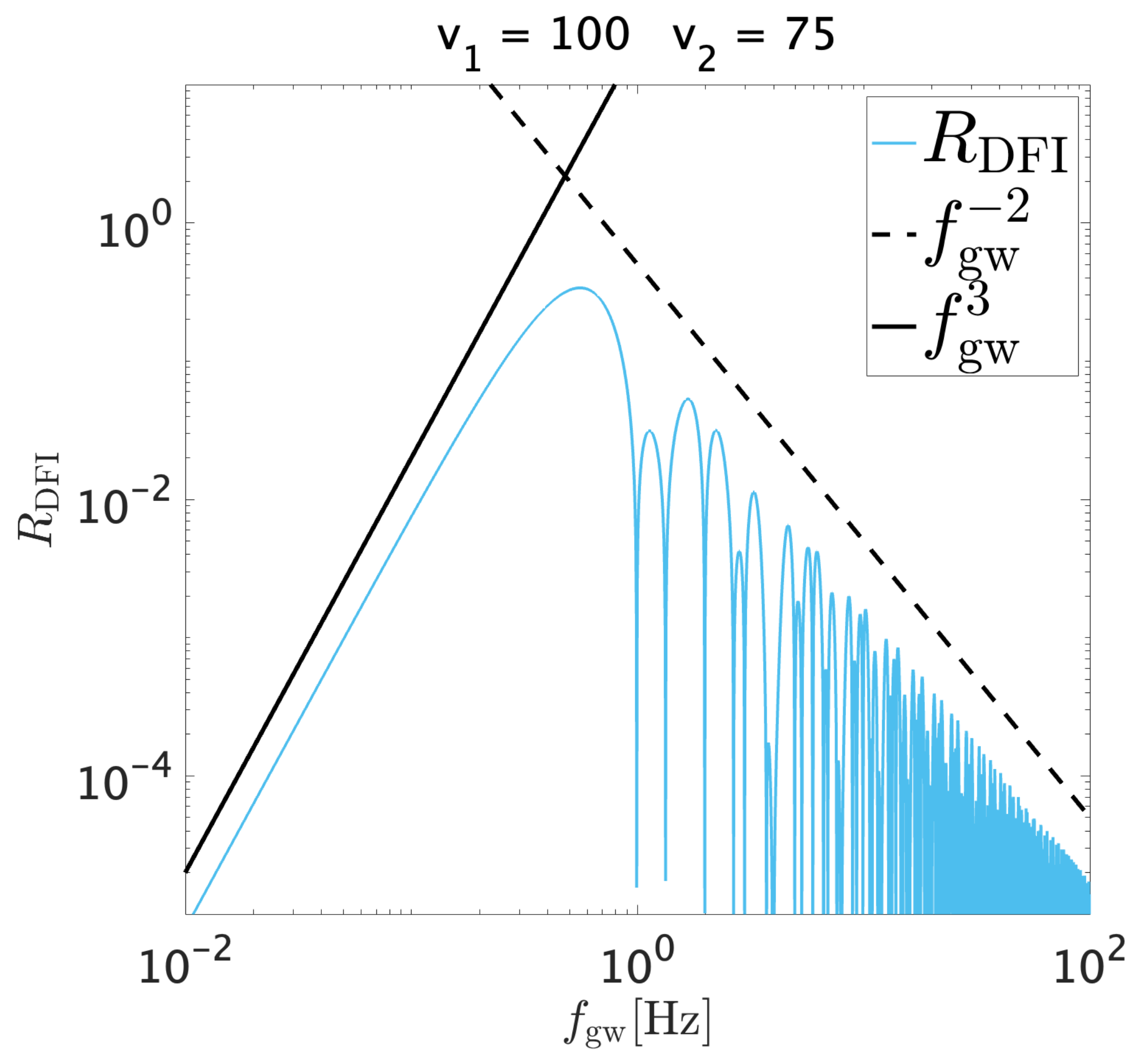}
    \end{minipage} &
    \begin{minipage}[b]{0.45\hsize}
      \centering
      \includegraphics[clip,width=7cm]{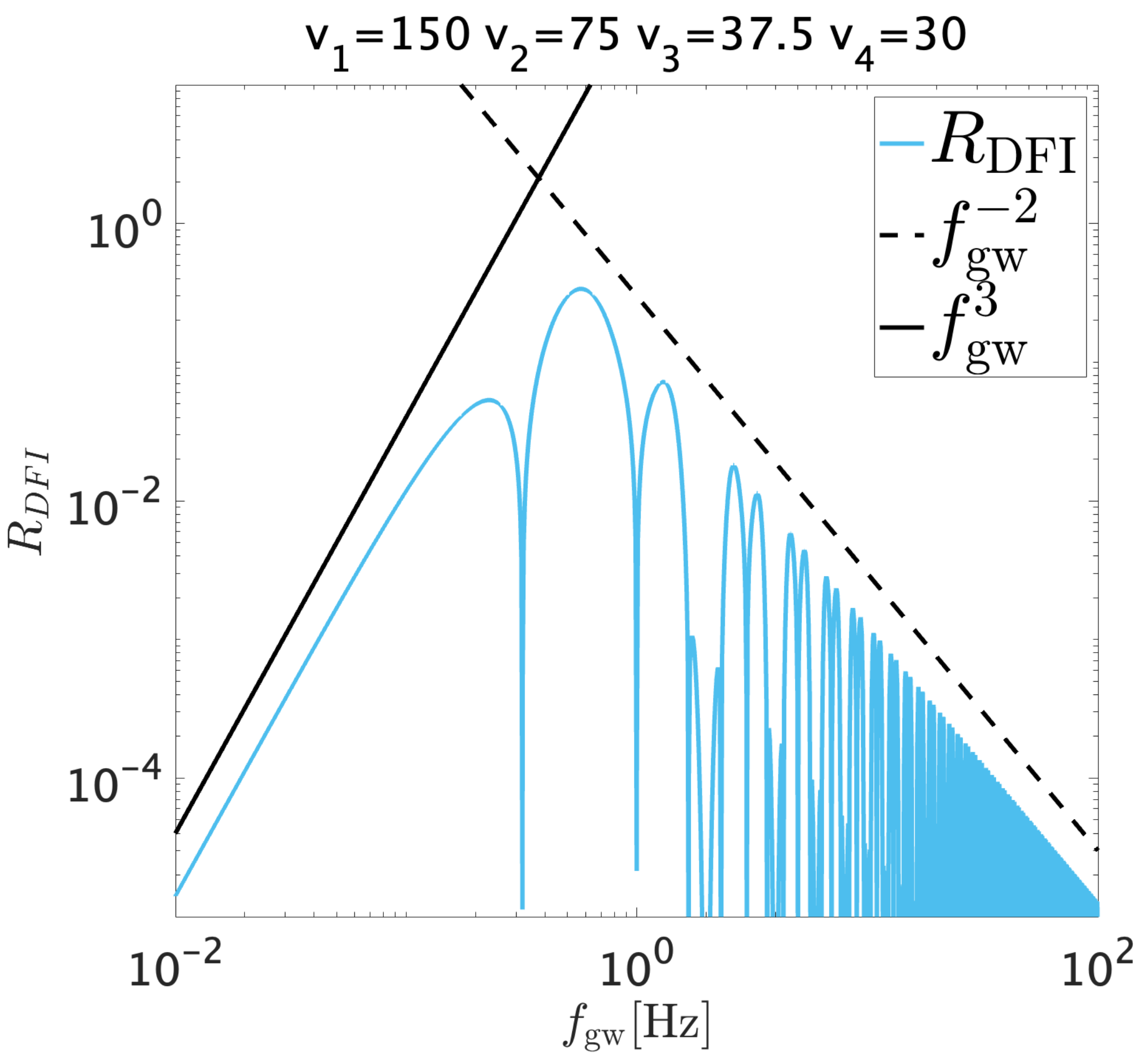}
    \end{minipage}
 \end{tabular}
 \caption{Response function $R_\mathrm{DFI} (\Omega)$ to GWs with the polarization of the cross mode ($\Psi = \pi / 4$). The diagonal solid and dashed lines are proportional to $f_\mathrm{gw}^3$ and $f_\mathrm{gw}^{-2}$. The left curve indicates the response function in the previous research \cite{DFNI_Iwaguchi}. The right curve indicates the one in this paper.}
 \label{fig:4}
\end{figure}

\noindent
Here, $\bar{\kappa}$ is defined as $\bar{\kappa} \equiv (\kappa_1+\kappa_2+\kappa_3+\kappa_4)/4$. When a GW with the polarization of the cross mode ($\Psi = \pi / 4$) propagates along the z axis ($\theta = 0$, $\phi = 0$), the response function to GWs of a single MZI with four unidirectional neutrons is shown in Fig.\ref{fig:4}. For $L=75.0$ m, $v_1=150$ m/s, $v_2=75.0$ m/s, $v_3=37.5$ m/s, $v_4=30.0$ m/s, and $\alpha=\pi/4$ rad, the neutron propagation times derived from Eq.(\ref{eq:15}) are $T_1=0.50$ s, $T_2=1.0$ s, $T_3=2.0$ s, and $T_4=2.5$ s. These times satisfy the condition for cancellation of beam splitter noise, that is, $T_1 + T_4 = T_2 + T_3$.

In the right panel of Fig.\ref{fig:4}, the GW response function for the neutron DFI is proportional to $f_\mathrm{gw}^3$ at lower frequencies and has a peak around 1 Hz. The GW response function at higher frequencies is proportional to $f_\mathrm{gw}^{-2}$. In the left panel, for $L=75.0$ m, $v_1=100$ m/s and $v_2=75.0$ m/s, the neutron propagation times derived from Eq.(\ref{eq:13}) are $T_1=0.75$ s and $T_2=1.0$ s. At the peak, around 1 Hz, the GW response function shown in the right panel is comparable to that in the left panel. Accordingly, the GW response function in this configuration has the same frequency dependence as the previous researches with the bidirectional neutrons. As a result, by maintaining the frequency dependence of the GW response function, it is possible to simplify the neutron DFI configuration in terms of the neutron incidence direction.

\section{Discussion \& Conclusions}
\label{sec:4}

We focused on the simplification for the neutron incidence direction and confirmed the feasibility of noise cancellation while maintaining GW signals in this simplified neutron DFI configuration, which consists of a single MZI with unidirectional injection of neutrons with different speeds. The cancellation of mirror displacement noise is realized by the combination of interferometric signals resulting from different-speed neutrons. The signals are normalized by the neutrons' speeds because the displacement noise of mirrors and beam splitters depends on the neutron speed. Here, the time of each signal is adjusted in such a way that neutrons impact the mirrors at the same time. The cancellation of beam splitter displacement noise depends on the frequency-dependent coefficients. These coefficients equalize the residual noise amplitudes and the specific condition for the neutrons' speeds. In a phasor diagram, this condition plays an essential role in making the displacement noise arrows parallel. The GW response function in this configuration shows the same frequency dependence as the configurations in previous research about the neutron DFIs.

It should be noted that the neutron DFI configuration described in this paper can be realized only in the absence of gravity. In the presence of gravity, neutrons with different speeds impact the mirrors at different locations if they hit the same point on the beam splitters. This difference occurs due to the parabolic trajectories of the neutrons, which are caused by gravity. Mirror displacement noise caused by longitudinal motion of the mirrors can be removed in the DFI scheme even though the neutrons do not hit the same point on the mirrors. However, the DFI cannot remove displacement noise caused by rotational motion of the mirrors because of its dependence on the impact points of neutrons. The difference in the rotational components of the displacement noise between two impact points increases with their separation. The internal thermal noise of the mirror also cannot be canceled because this noise depends on the impact point. Note that the neutron DFI schemes in the previous research \cite{DFNI_Iwaguchi} do not have this problem. In those configurations, the displacement noise of the mirrors at each point can be canceled by combining two interferometer signals from neutrons propagating in opposite directions and at the same speed, because they hit the same points on the mirrors. On the contrary, in the configuration described in this paper, neutrons with different speeds do not hit the same points on the mirrors.

This problem is not critical for a proof-of-principle experiment of a neutron DFI. This is because a small experimental setup, where the distance between the impact points of neutrons on the mirrors will be negligible due to the short propagation time of neutrons, suffices to demonstrate the principle of the neutron DFI. We plan to carry out a proof-of-principle experiment for the neutron DFI with this unidirectional configuration in the near future.

\section*{Acknowledgment}
We would like to thank R. L. Savage for English editing. This work was supported by the Japan Society for the Promotion of Science (JSPS) KAKENHI Grant Number JP19K21875. A. N. is supported by JSPS KAKENHI Grant Nos. JP19H01894 and JP20H04726 and by Research Grants from the Inamori Foundation.

\end{document}